\documentstyle[12pt]{article}
\begin{document}

\def\slash{\!\!\!\!/\,}
\def\dirac{{D}\slash}
\def\ham{{\bf \rm H}}
\def\wilson{{\bf \rm B}}
\def\chiral{{C}\slash}

\title{ 
\vspace{-5cm}
\begin{flushright}
{\normalsize FSU-SCRI-97-112}\\
\end{flushright}
\vspace{2.5cm}
A new method to measure the chiral condensate 
in lattice QCD using Wilson fermions}
\author{
Khalil M.~Bitar \\
SCRI, The Florida State University, \\
Tallahassee, FL 32306-4130, USA
\\ and \\
American University of Beirut, Lebanon
\\
\\
Urs M. Heller and Rajamani Narayanan
\\
SCRI, The Florida State University, \\
Tallahassee, FL 32306-4130, USA
}

\maketitle

\begin{abstract}
Parity and flavor symmetry is not broken in QCD. Using this fact,
we propose a new method to measure the chiral condensate in lattice
QCD using Wilson fermions with an operator that breaks parity and flavor
in addition to the chiral symmetry.
\end{abstract}

\noindent
{\bf PACS \#:}  12.38.-t, 11.30.Rd, 11.15.Ha.\hfill\break

\section{Introduction}

Formal arguments in the continuum have shown that parity and flavor
symmetries are not broken in QCD\cite{vwp,vwf}. On the other
hand chiral symmetry is expected to be spontaneously broken in massless
QCD with pions being the associated Goldstone bosons. A non-vanishing density
of eigenvalues around zero for the massless Dirac operator 
is responsible for the breaking of the chiral symmetry\cite{bc}.
A promising way to test the above statements in a rigorous non-perturbative
setting is the lattice formulation of QCD. Attempts have been made to
study these issues in lattice QCD using Wilson fermions and staggered fermions.
For Wilson fermions chiral symmetry is broken explicitly by the Wilson
term needed to remove the unwanted doublers on the lattice. 
This makes
it difficult to study the spontaneous breaking of chiral symmetry using
Wilson fermions since
there is an additive renormalization to the  bare fermion
bilinear used to measure the chiral condensate on the lattice.
Staggered fermions do realize massless fermions on the
lattice but the continuum notion of 
chirality is not well defined for staggered fermions due to
their inherent lattice construction. 

Although parity and flavor symmetry is not
 broken in the continuum theory there are claims that it
could be broken on the lattice in the context of Wilson fermions\cite{Aoki}.
If this statement extends to the continuum limit, then it is not clear
if Wilson fermions do realize QCD. A recent numerical investigation has
provided some evidence that parity and flavor symmetry is probably not broken by
Wilson fermions as one approaches the continuum
limit\cite{Bitar}. In this letter we show that parity and flavor symmetry is not
broken by Wilson fermions on the lattice even away from the continuum limit.
As a consequence we propose a parity-flavor breaking operator as
a good way to probe the spontaneous breaking of chiral symmetry using
Wilson fermions on the lattice.

We first discuss the issue of parity and flavor symmetry in
the continuum. We emphasize that the result for the preservation of
 flavor symmetry in~\cite{vwf} is only valid for
the massive theory although it can be extended to the massless case
with some care. As for parity,
we note that one can rotate a $m\bar\psi\psi$ term into an
$ih\bar\psi\gamma_5\psi$ term by a chiral rotation. 
One has  to take the limit $h\rightarrow 0$
for a finite $m$ to study the violation of parity in massive QCD.
On the other hand, one can
approach the $m=h=0$ limit
in the complex $(m,h)$ plane in any direction to measure
the chiral condensate. 
With the above continuum statements in place, we focus on the lattice formulation
of two flavor QCD using Wilson fermions. By introducing a parity and flavor
breaking term, we show that parity and flavor symmetry is not broken in the massive
theory realized using Wilson fermions. In the massless theory, statements
similar to the continuum hold and this leads us to probe the chiral condenstate
using an operator that breaks parity and flavor symmetry with no need for 
subtractions. 

\section{Continuum}

The fermionic part of the continuum Euclidean action for QCD with 
explicit parity, flavor and chiral breaking terms 
is
\begin{equation}
S = \sum_f \Bigl[ \bar\psi_f \dirac \psi_f + m_f \bar\psi_f\psi_f 
+ i h_f \bar\psi_f\gamma_5 \psi_f \Bigr].
\label{eq:cont_action}
\end{equation}
The subscript $f$ stands for the flavor index and 
$\dirac=\gamma_\mu D_\mu$ is the
usual Euclidean massless Dirac operator in a background gauge field .
$D_\mu(A)$, $\mu=1,2,3,4$ are the four covariant derivative operators
and they are anti-Hermitean. All the $\gamma_\mu$'s are Hermitean and therefore
$\dirac$ is an anti-Hermitean operator. Note that $\bar\psi_f$ and $\psi_f$ are
independent degrees of freedom in Euclidean space. We recall that
$\{\gamma_5, \dirac\}=0$ and that $\gamma_5\dirac$ is a Hermitean
operator. Under a chiral rotation given by
\begin{equation}
\psi_f \rightarrow e^{i\theta\gamma_5} \psi_f;\ \ \ \ 
\bar\psi_f \rightarrow \bar\psi_f e^{i\theta\gamma_5} ,
\label{eq:chiral_trans}
\end{equation}
Eqn.(\ref{eq:cont_action})
goes into itself with
\begin{equation}
(m_f,h_f) \rightarrow 
(m_f\cos(2\theta)-h_f\sin(2\theta), h_f\cos(2\theta) + m_f\sin(2\theta))
\label{eq:chiral_rot}
\end{equation}
for each of the flavors. On the other hand,
under a parity transformation, $(m_f,h_f) \rightarrow (m_f, -h_f)$.
A parity tranformation can be realized as a chiral rotation in the massless
case and therefore some care has to be taken in dealing with the massless
theory.
We assume that the theory is formulated in a periodic box of finite
volume and we will then take the
limit of the volume going to infinity in order to
study spontaneous breaking of symmetries. The spectrum is discrete in 
a finite
volume and is given by
\begin{equation}
\dirac e_n = i\lambda_n e_n; \ \ \ \ 
\dirac \gamma_5 e_n = -i\lambda_n \gamma_5 e_n;\ \ \ \ 
\lambda_n > 0 ,
\label{eq:eigen}
\end{equation}
and the eigenfunctions form a complete orthnormal set, i.e,
\begin{equation}
e^\dagger_n e_m = \delta_{mn};\ \ \ \ 
e^\dagger_n \gamma_5 e_m = 0 .
\end{equation}
We can expand $\psi_f$ and $\bar\psi_f$ in this orthonormal basis as follows:
\begin{equation}
\psi_f = \sum_n (\phi_f^n e_n + \chi_f^n \gamma_5 e_n);\ \ \ \ 
\bar\psi_f = \sum_n 
(\bar\phi_f^n e^\dagger_n + \bar\chi_f^n e^\dagger_n \gamma_5).
\label{eq:expand}
\end{equation}
where $\phi_f^n$, $\chi_f^n$, $\bar\phi_f^n$ and $\bar\chi_f^n$ are the
new Grassmann variables in the eigenfunction basis and the change of basis
has a Jacobian equal to unity. 
In this basis, the action decouples into various sectors with each sector
associated with $e_n,\gamma_5 e_n$. Eqn.(\ref{eq:cont_action}) can be rewritten
as
\begin{equation}
S = \sum_f \sum_n 
\Bigl [ i\lambda_n \bar\phi^n_f \phi^n_f - i\lambda_n \bar\chi^n_f\chi^n_f 
       + m_f (\bar\phi^n_f \phi^n_f + \bar\chi^n_f\chi^n_f) 
       + ih_f (\bar\phi^n_f \chi^n_f + \bar\chi^n_f\phi^n_f)  \Bigr ].
\label{eq:action_modes}
\end{equation}
With this in place, one can write expressions for the two fermionic
expectation values of interest in each flavor sector, namely,
\begin{equation}
\langle \bar\psi_f \psi_f \rangle_A = \sum_n {2 m_f \over \lambda_n^2 + m_f^2 + h_f^2},
\label{eq:chiral}
\end{equation}
\begin{equation}
i\langle \bar\psi_f \gamma_5 \psi_f \rangle_A 
= \sum_n {2 h_f \over \lambda_n^2 + m_f^2 + h_f^2},
\label{eq:parity}
\end{equation}
where both of the expectation values above are in a fixed gauge field background.
$m_f^2 + h_f^2$ is invariant under the chiral rotation given by
Eqn.(\ref{eq:chiral_rot})
and only this combination appears in the denominator
of the above two equations.
In the infinite volume limit, the sums in Eqns. (\ref{eq:chiral}) and 
(\ref{eq:parity}) are replaced by the integrals,
\begin{equation}
\langle \bar\psi_f \psi_f \rangle_A = {m_f\over \sqrt{m_f^2 + h_f^2}} 
\int_{-\infty}^{\infty} d\lambda \rho(\lambda){1 \over \sqrt{(m_f^2 + h_f^2)} + i\lambda},
\label{eq:chiral_cont}
\end{equation}
\begin{equation}
i\langle \bar\psi_f \gamma_5 \psi_f \rangle_A = {h_f\over \sqrt{m_f^2 + h_f^2}} 
\int_{-\infty}^{\infty} d\lambda \rho(\lambda){1 \over \sqrt{(m_f^2 + h_f^2)} + i\lambda},
\label{eq:parity_cont}
\end{equation}
with $\rho(\lambda)d\lambda$ being the density of eigenvalues at $\lambda$.
From Eqn.(\ref{eq:eigen}) it follows that $\rho(\lambda)$ is an even function of
$\lambda$. 

Vafa and Witten~\cite{vwf} showed that flavor symmetry is not broken in massive QCD
by first noting that in the limit of infinite volume 
\begin{equation}
\langle \bar\psi_1 \psi_1 - \bar\psi_2 \psi_2 \rangle_A =
\int_{-\infty}^{\infty} d\lambda \rho(\lambda) \Bigl (
{1\over m_1 + i\lambda} -{1\over m_2 + i\lambda} \Bigr ) .
\label{eq:flavor_break}
\end{equation}
We have set $h_1=h_2=0$ in Eqn.(\ref{eq:chiral_cont}) so that parity is not explicitly broken. 
Since $m_1\ne 0$ and $m_2\ne 0$ there is no singularity in the integration
region and therefore each of the integrals separately is finite. In the limit
of $m_1=m_2$ they will be the same and we will arrive at 
$\langle \bar\psi_1 \psi_1 - \bar\psi_2 \psi_2 \rangle_A=0$ for every gauge
field background indicating that flavor symmetry is not broken in massive QCD.

A similar argument also shows that parity is not broken in massive QCD. 
Since $m_f\ne 0$, there is no singularity in the integration region of
Eqn.(\ref{eq:parity_cont}) with $h_f=0$ and we have
\begin{equation}
\lim_{h_f\rightarrow 0} i \langle \bar\psi_f\gamma_5 \psi_f \rangle_A|_{m_f\ne 0} = 0.
\label{eq:parity_conserve}
\end{equation}
This argument is different from the one presented in \cite{vwp}. 
In \cite{vwp}, the authors 
make the assumption that the free energy is a smooth function of the parity
breaking parameter $h$ and 
then show that parity cannot be broken by introducing
an arbitrary parity breaking operator. 
It is not apriori clear if the assumption
of the smoothness of the free energy as a function of $h$ is a valid one.
A spontaneous breaking of parity could result in a discontinuity in the slope of
the free energy at $h=0$. Here we have considered a fermionic operator that breaks 
parity and have shown that its expectation value is zero. 

Care has to be taken in extending the arguments above to massless QCD. 
Following~\cite{bc} we note that
\begin{equation}
\lim_{\mu\rightarrow 0} \int d\lambda \rho(\lambda) {i\over \lambda \pm i\mu} =
\pm \pi \rho(0) + i \int d\lambda \rho(\lambda) P({1\over\lambda}),
\label{eq:bc}
\end{equation}
where $P({1\over\lambda})$ denotes the principle value of ${1\over\lambda}$ and
is an odd function of $\lambda$. The first term arises from the singularity at $\lambda=0$
in the limit of $\mu=0$. 
If we set $m_f=m$ and $h_f=h$, we have
\begin{equation}
\lim_{h\rightarrow 0^+} i\langle \bar\psi \gamma_5 \psi \rangle_A|_{m=0} = 
\lim_{m\rightarrow 0^+} \langle \bar\psi \psi \rangle_A|_{h=0} = \pi \rho(0)
\label{eq:parity_chiral}
\end{equation}
where we have used Eqn.({\ref{eq:bc}) with $\mu=m$ and $\mu=h$. 
If $\rho(0)\ne 0$ we have a spontaneous breakdown of
chiral symmetry. In fact one can approach the chiral limit $(m=h=0)$ along
any direction in the complex $(m,h)$ plane and measure the magnitude of the chiral
condensate. 

To study flavor symmetry breaking in massless QCD, we proceed
as follows. We set $h_f=0$ so that the only term that breaks
flavor symmetry are the mass terms. In the massless limit,
\begin{equation}
\lim_{m_f\rightarrow 0} 
\langle (\bar\psi_f \psi_f  \rangle_A|_{h=0}
= \pi {m_f\over |m_f|} \rho(0) 
\label{eq:massless}
\end{equation}
and this will be non-zero 
since chiral symmetry is expected to be broken
in the massless limit.
Clearly,
\begin{equation}
\lim_{m_1\rightarrow 0} 
\Bigl| \langle (\bar\psi_1 \psi_1  \rangle_A|_{h=0}\Bigr|
= 
\lim_{m_2\rightarrow 0} 
\Bigl|\langle (\bar\psi_2 \psi_2  \rangle_A|_{h=0}\Bigr|
\label{eq:flavor_conserve}
\end{equation}
and therefore the flavor symmetry is not broken.
Note that
\begin{equation}
\lim_{m_1\rightarrow 0^+} 
 \langle (\bar\psi_1 \psi_1  \rangle_A|_{h=0}
= 
- \lim_{m_2\rightarrow 0^-} 
\langle (\bar\psi_2 \psi_2  \rangle_A|_{h=0}
\label{eq:flavor_break_massless}
\end{equation}
But this is
just another way to see the breaking of chiral symmetry.
A similar statement holds if we study flavor symmetry 
breaking by setting $m_f=0$ and approach $h_f\rightarrow 0$. 

We close this section by remarking that the eigenvalues of 
$\ham(m)= \gamma_5(\dirac+m) $ are
$\pm \sqrt{\lambda_n^2 + m^2}  $ with the same
$\lambda_n$'s as in Eqn.(\ref{eq:eigen}). The spectrum is real
and has a gap when $m\ne 0$. It is natural to consider $\ham(m)$ in the context
of Wilson fermions on the lattice.

\section{Lattice}

We consider the issue of spontaneous breaking of chiral, parity and flavor symmetry
on the lattice in the context of Wilson fermions. The lattice equivalent of
$\ham(m)$ is
\begin{equation}
\ham_L(m) = \pmatrix{ \wilson - m & \chiral \cr
\chiral^\dagger & -\wilson +m  \cr}
\label{eq:latham}
\end{equation}
with $\chiral = \sum_\mu\sigma_\mu C_\mu$ and $\wilson = \sum_\mu\wilson_\mu$.
$\sigma_1$, $\sigma_2$, $\sigma_3$ are the usual Pauli matrices and $\sigma_4 = iI$. 
$C_\mu$ and $\wilson_\mu$ are the symmetric
first covariant derivative and second
covariant derivative on the lattice respectively. $m$ is the bare lattice mass.
The $\wilson$ part of $\ham_L(m)$ is $\gamma_5$ times the usual Wilson term.
The lattice action is 
\begin{equation}
\sum_f \psi_f^\prime H_L(m) \psi_f + ih_f \psi_f^\prime \psi_f.
\label{eq:lattice_action}
\end{equation}
$\psi_f^\prime$ is related to the conventional $\bar\psi_f$ by 
$\bar\psi_f=\psi_f^\prime\gamma_5$.
Following \cite{Aoki,Bitar}, we have introduced a parity breaking term equal to
 $ih_f \psi_f^\prime \psi_f$. 
Since both flavors couple to the same gauge field and have the same lattice mass,
the only
flavor symmetry breaking term arises from the different couplings to the parity
breaking term. 
In particular we will 
consider two flavor QCD with $h_1=-h_2=h$ so that we have one coupling that
breaks both parity and flavor
symmetry. We can then study the theory in the limit of $h\rightarrow 0$.
In addtion to $h$ we have the lattice gauge coupling, $\beta$
and the lattice mass $m$. 
The functional integral 
is positive definite for all values of $m,h$ and $\beta$. The gauge action
is positive as always and the result of the fermion integral is 
$\det (\ham_L^2 + h^2)$
which is positive definite. 

Like in the continuum,
the expectation value of the parity-flavor breaking term in a fixed gauge
field background on an infinite lattice is 
\begin{equation}
\langle i(\bar\psi_1\gamma_5 \psi_1 - \bar\psi_2\gamma_5 \psi_2) \rangle_A =
\int_{-\infty}^{\infty} d\lambda \rho(\lambda)
\Bigl [{i\over \lambda + ih} - {i\over \lambda - ih}\Bigr] .
\label{eq:latparflav}
\end{equation}
Here the 
$\lambda$'s are the eigenvalues of $\ham_L(m)$ on the lattice. The eigenvalues
can be positive and negative but are bounded. In the continuum limit $\rho(\lambda)$
will be an even function of $\lambda$ but this will not be the case for finite
$\beta$. 
If the spectrum has a gap, the integral is well defined and finite and therefore
$\lim_{h\rightarrow 0} 
\langle i(\bar\psi_1\gamma_5 \psi_1 - \bar\psi_2\gamma_5 \psi_2) \rangle_A
=0$ and there is no breaking of parity or flavor symmetry. This is indeed the case
if $m < m_c(\beta)$ where the lattice theory describes massive fermions.
If the spectrum does not have a gap, then 
inserting Eqn.(\ref{eq:bc}) into
Eqn.(\ref{eq:latparflav}) yields
\begin{equation}
\lim_{h\rightarrow 0}
\langle i(\bar\psi_1\gamma_5 \psi_1 - \bar\psi_2\gamma_5 \psi_2) \rangle_A
= 2\pi \rho(0)
\label{eq:main_result}
\end{equation}
This shows that a non-zero result for the parity-flavor breaking operator on
an infinite lattice at any finite value of $\beta$ arises from a non-vanishing
spectral density of $\ham_L(m)$ at zero. As remarked in the continuum analysis
this is simply a signal for the spontaneous breakdown of chiral symmetry.
In particular if we had set the parity breaking term for both flavors to be
of the same sign and then considered Eqn.(\ref{eq:latparflav}), we would have found
that the $\rho(0)$ contribution from the two flavors cancel and we would have
obtained a zero result indicating that flavor symmetry is not broken just
like in the massive theory. 

In the continuum limit,
we expect the spectrum to have a gap if $m\ne 0$ and the
gap to close at $m=0$. At a finite $\beta$ the
$\wilson$ term in Eqn.(\ref{eq:latham}) 
plays a role and the effective mass of the fermion is gauge field
dependent. As such there cannot be a precise value of $m$ where the gap closes.
This is evident when one studies topological issues using $\ham_L(m)$~\cite{over_top}.
For a topologically non-trivial configuration on the lattice, $\ham_L(m)$ must have
a zero eigenvalue for some value of $m$~\cite{over}. The value of $m$ where one
gets a zero eigenvalue is a function of the size of the topological object.
If we take the point of view that a non-zero $\rho(0)$ is a consequence of the
gauge field configuration containing  instantons and anti-instantons, we
 expect $\rho(0)$ to be non-zero for a range of $m$ if the lattice coupling is finite.
Even if we do not accept the above point of view it is likely that there is a region of
$m$ where $\rho(0)$ is non-zero since there is no precise definition of a massless point
at a finite $\beta$. 
For Wilson fermions, the region of $m$ where $\rho(0)$ is non-zero is of order of the
lattice spacing and will thus shrink to a point at $m=0$ in the continuum limit.
$\lim_{h\rightarrow 0}
\langle i(\bar\psi_1\gamma_5 \psi_1 - \bar\psi_2\gamma_5 \psi_2) \rangle_A$
will be non-zero over a region of $m$ on the lattice at finite $\beta$. 
In this whole region of $m$ we are simply measuring a breakdown of chiral symmetry
and we could say that a measure of the total chiral symmetry on the lattice is
to sum the condensate over the whole region of $m$.
This argument leads us to the following proposal for the measurement of the chiral
condensate using Wilson fermions on the lattice:
\begin{equation}
<\bar\psi\psi> ={1\over 2} \lim_{\beta\rightarrow\infty}
{1\over m_2(\beta)-m_1(\beta)}\int_{m_1(\beta)}^{m_2(\beta)} dm 
\lim_{h\rightarrow 0} 
\lim_{V\rightarrow\infty} 
\langle i(\bar\psi_1\gamma_5 \psi_1 - \bar\psi_2\gamma_5 \psi_2) \rangle
\label{eq:proposal}
\end{equation}
The order of the limits are standard. We first take the infinite volume limit
before taking the coefficient of the parity-flavor breaking term to zero.
For finite $\beta$ we will get a non-zero result if $m_1(\beta) \le m \le
m_2(\beta)$ justifying the integral over this region of $m$. 
The continuum limit is taken at the very end. 
In the continuum limit both $m_1(\beta)$ and $m_2(\beta)$ approach zero
and the integral reduces to the value of the integrand at $m=0$. 
We note that the second term in Eqn.(\ref{eq:bc}) contributes to 
$\langle i\bar\psi_i\gamma_5\psi_i \rangle_A$ in Eqn.(\ref{eq:latparflav}) for a finite
$\beta$ since $\rho(\lambda)$ is not an even function of $\lambda$.
But this contribution is zero in the continuum limit and is exactly cancelled between
the two terms in Eqn.(\ref{eq:proposal}) on the lattice.

\section{Conclusions}

We have considered the issue of parity and flavor symmetry breaking in QCD. By an
extension of the standard analysis in the continuum~\cite{bc,vwf} 
we have shown that parity and flavor symmetry is not broken in massless
QCD. Similar arguments show that parity and flavor symmetry is not broken in the
lattice formulation of QCD using Wilson fermions. This lead us to propose a
measurement of the
breakdown of chiral symmetry using an operator that breaks parity and flavor symmetry
in lattice QCD with Wilson fermions. 
An advantage of this proposal is that the operator does not suffer from any
additive renormalization. Like in any numerical study of a spontaneous breakdown
of a symmetry, it should be possible to use results
on several finite $V$'s and several non-zero $h$'s and estimate the limit as
$V\rightarrow\infty$ and $h\rightarrow 0$ by standard extrapolations. 

\vskip 1cm

\noindent {\bf Acknowledgements:} We would like to thank Robert Edwards and
Ivan Horvath for useful discussions. This research was supported by DOE contracts 
DE-FG05-85ER250000 and DE-FG05-96ER40979.

\end{document}